\documentclass{aa}
\usepackage{graphicx}
\usepackage{hyperref}
\usepackage{txfonts}
\usepackage{latexsym}
\usepackage{supertabular,multirow}

\begin{document}

\title{Diffuse Interstellar Bands in NGC 1448\thanks{Based 
on observations collected at the European Southern Observatory, Paranal, Chile
(ESO Programmes 67.D-0227 and 71.D-0033).}
}

\author{
Jesper Sollerman,\inst{1}
Nick Cox,\inst{2}
Seppo Mattila,\inst{1}
Pascale Ehrenfreund,\inst{2,3}
Lex Kaper,\inst{2}
Bruno Leibundgut,\inst{4}
\and
Peter Lundqvist\inst{1}
}

\institute{
Stockholm Observatory, Department of Astronomy, AlbaNova, 
SE-106 91 Stockholm, Sweden  
\and 
Astronomical Institute "Anton Pannekoek", University of Amsterdam, 
Kruislaan 403, NL-1098, Netherlands
\and
Astrobiology Laboratory, Leiden Institute of Chemistry, P.O.Box 9502, 
2300 RA Leiden, Netherlands
\and
European Southern Observatory, Karl-Schwarzschild-Strasse 2, Garching, 
D-85748, Germany
}

\date{Received --- ?, Accepted --- ?} 
\authorrunning{Sollerman et al.}
\titlerunning{DIBs in NGC 1448} 
\offprints{Jesper Sollerman;\hfill\\
e-mail: jesper@astro.su.se}

\abstract{ 
We present spectroscopic VLT/UVES observations of two emerging supernovae,
the Type Ia SN\,2001el and the Type II SN\,2003hn, in the spiral galaxy
NGC\,1448.  Our high resolution and high signal-to-noise spectra
display atomic lines of Ca\,{\sc ii},  Na\,{\sc i}, Ti\,{\sc ii} and
K\,{\sc i} in the host galaxy.
In the line of sight towards SN\,2001el, we also detect over a dozen
diffuse interstellar bands (DIBs) within NGC\,1448. 
These DIBs have 
strengths comparable to
low reddening galactic lines of sight, albeit with some variations.  
In particular, a good match is found with the line of sight towards the
$\sigma$ type diffuse cloud (HD\,144217).
The DIBs towards SN\,2003hn are significantly
weaker, and this line of sight has also lower sodium column density.
The DIB central velocities show
that the DIBs towards SN\,2001el are closely related to the 
strongest  interstellar Ca\,{\sc ii} and Na\,{\sc i} components, 
indicating that the DIBs are preferentially produced in the same cloud.
The ratio of the 
$\lambda$\,5797 and $\lambda$\,5780 DIB strengths (r $\sim$ 0.14) suggests
a rather high UV field in the DIB environment towards SN\,2001el.
We also note that the extinction estimates obtained from the sodium lines
using multiple line fitting agree with reddening estimates based on 
the colors of the Type Ia SN\,2001el.

}

\maketitle

\section{Introduction}

\subsection{Extragalactic DIBs}

The Diffuse Interstellar Bands (DIBs) are a large number of absorption
lines between $\sim4000 - 10\,000$~\AA~that are superimposed on the
interstellar extinction curve (e.g., Herbig 1995). 
During the last 7 decades of DIB studies
almost 300 DIBs have been detected.
Within the Milky Way, DIBs have been observed towards more than
a hundred stars. However, there is still no definitive identification
of the DIB carriers. 
Recent studies indicate that the environmental behaviors
of DIBs reflect an interplay between ionization, recombination,
dehydrogenation and destruction of chemically stable species 
(\cite{Herbig}; \cite{C97}; \cite{VF00}).
It is therefore of interest to study DIBs in different environments, 
especially in external galaxies.

Hitherto, only a handful of DIBs have 
been observed in extragalactic targets
(e.g., \cite{V87}; \cite{M87}; \cite{HL00}; \cite{E02}).
The Magellanic Clouds have been most intensely studied in this
respect. Detailed views of LMC DIBs were obtained towards the bright
supernova (SN) 1987A (\cite{V87}). Today, high
resolution spectra can also be obtained of reddened stars in the LMC
with large telescopes (Ehrenfreund et al. 2002). 
For more distant galaxies, however, supernovae still provide the most
promising opportunity to probe the extragalactic interstellar
medium. Spectra taken of SN\,1986G in the nearby galaxy NGC 5128 
(\cite{Rich87}; \cite{DO89}) allowed the detection of a 
few extragalactic DIBs outside the Local Group. 
Some DIBs were also tentatively 
detected towards SN\,1989M in NGC 4579 (\cite{Steidel90}).

In this paper we present high-resolution
observations of two emerging supernovae in NGC~1448. The data of the 
well studied Type Ia SN\,2001el allowed us to detect
more than a dozen extragalactic DIBs with unprecedented
signal-to-noise. At a different line-of-sight through the same galaxy, 
the Type II SN~2003hn did not show the same spectacular DIB signal.

\subsection{SNe 2001el and 2003hn in NGC 1448}

\begin{figure*}[t]
\begin{center}
\includegraphics[width=90mm, angle=0, clip] {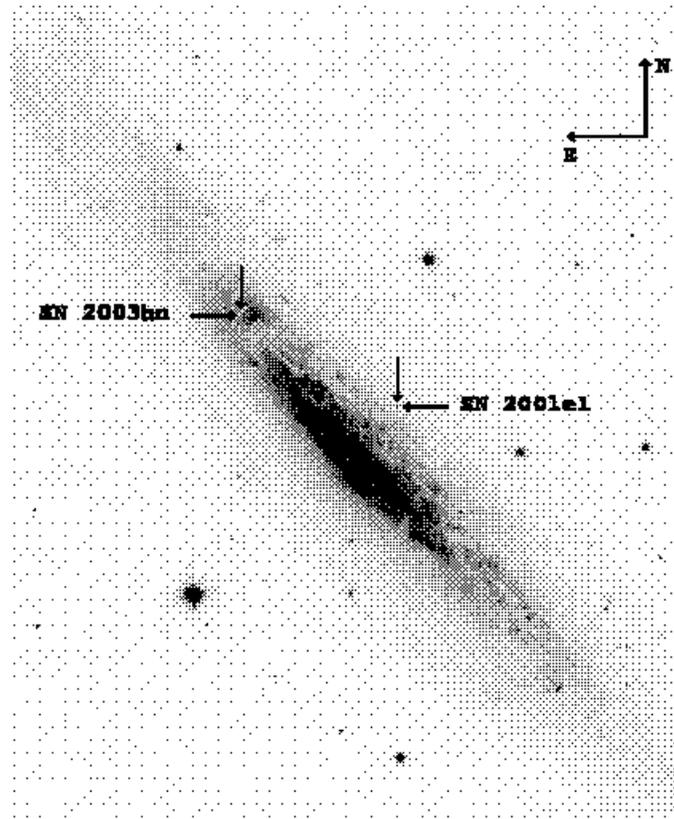}
\end{center} 
 \caption{Supernova 2001el as observed in the V-band with FORS1 on VLT in
August 2002, i.e., almost one year past explosion. This late image shows
the position of the supernova within the galaxy, NGC 1448. The field of view
of the image is $4\farcm5\times5\farcm1$. North is up and East to the
left. The locations of the two supernovae, SNe 2001el and 2003hn, are marked
by arrows.}\label{fig1}
\end{figure*}

Supernova 2001el was discovered on September 17.1 (UT) 2001 (\cite{Monard01}).
It was situated about 14\arcsec~West and 20\arcsec~North of the
nucleus of the nearby warped spiral 
galaxy NGC 1448 (Fig.~\ref{fig1}).
Within our 
Target-of-Opportunity programme to carry out 
early high resolution spectroscopy of nearby supernovae 
(e.g., \cite{Lundqvist04}),
we obtained a first spectrum on September 21. This
allowed a classification of the supernova as a Type Ia observed well
before maximum (\cite{Sollerman01}).

SN\,2001el reached its maximum magnitude ($B=12.8 mag$) on 2001 September 30,
and thereby became the brightest  supernova
that year. This supernova has been well monitored both photometrically
and spectroscopically, and has been shown to be a normal Type Ia
supernova 
(\cite{K03}).

SN\,2003hn was discovered in the same galaxy 
on August 25.7 2003 (\cite{Evans03}). 
It was  located approximately 47\arcsec~East and 53\arcsec~North 
of the nucleus (Fig.~\ref{fig1}).  Spectroscopy showed this to be a
Type II supernova approximately 2 weeks after explosion (\cite{Salvo03}).  
In this case we were motivated
by our previous detection of 
DIBs  against SN\,2001el in
the very same galaxy. We therefore executed high-resolution  spectroscopy
also for SN\,2003hn, to probe  the interstellar matter in another
line-of-sight in NGC~1448.

In Sect. 2 we will outline the observations and data reduction 
procedures. The
results are  then presented in Sect. 3 and discussed in Sect. 4.
We summarize our conclusions in Sect. 5.

\section{Observations and data reduction}

All observations were obtained with the  Ultraviolet and Visual
Echelle Spectrograph (UVES)\footnote{www.eso.org/instruments/UVES/} 
on the second unit telescope (Kueyen) of the Very Large Telescope (VLT)
on Paranal, Chile.  UVES is a high-resolution two-arm cross-dispersed
Echelle spectrograph,  where both arms can be operated simultaneously
using a dichroic beamsplitter (\cite{Kaufer02}).  
This enables high efficiency
from the atmospheric cutoff in the blue to the  long-wavelength limit
of the CCDs in the red.

On the night of September 21, we obtained 4 exposures of 2400 seconds
each of SN\,2001el. These were divided into two setups, in order to
obtain a complete wavelength coverage. The log of all our observations
of SN\,2001el is given in Table~\ref{t:obs01el}.
The supernova was observed again on September 26 and was
revisited for the last time on September 28.
The observations were
thus obtained 9, 4 and 2 days before maximum light of the supernova.

The observations of SN\,2003hn were obtained on August 31, 2003. We
obtained 3 exposures of 1400 seconds each in both the red and blue
set-ups. The details are given in Table~\ref{t:obs03hn}.  
This supernova was only observed
once, and since it was also at least one magnitude fainter than 
SN\,2001el at the time of our observations, the signal-to-noise of the 
SN\,2003hn data is not as good as for SN\,2001el.

The spectra were interactively reduced using the
UVES-pipeline\footnote{www.eso.org/observing/dfo/quality/ 
(versions 1.4.0 and 2.0)}
as implemented in $\tt MIDAS$.
This reduction package
allows for bias subtraction and flat-fielding of the  data using
calibration frames obtained in the morning.  Wavelength calibration
can be very accurately achieved  by comparison to ThAr arc lamps.


In searching for the DIBs, we summed together all the observations
from  epochs 1 and 3 
for SN\,2001el, when applicable 
(see Table~\ref{t:obs01el}).  
The second epoch was not added to the final spectrum, since
the exposure time was shorter and the seeing was worse at this
epoch. 
For SN\,2003hn all available data were combined.

\section{Results}

\subsection{Interstellar atomic lines}

Superposed on the spectra of SNe\,2001el and 2003hn we detect
interstellar atomic absorption lines, both from the 
Milky Way (MW; $l=251.5$, $b=-51.4$) 
and from NGC\,1448. The detected lines are Ca\,II K\&H (3933.66,
3968.47~\AA), Na\,I D (5889.95, 5895.92~\AA) and Ti\,II
(3383.76~\AA).  K\,I (7664.90, 7698.96~\AA) was also detected
for SN\,2001el, albeit with much weaker signal. 
The strongest components of these lines come from absorption within
NGC\,1448 (see Table~\ref{tb:ISfit}).

\begin{figure*}[t]
\begin{center}
\includegraphics[width=130mm, angle=-90, clip] {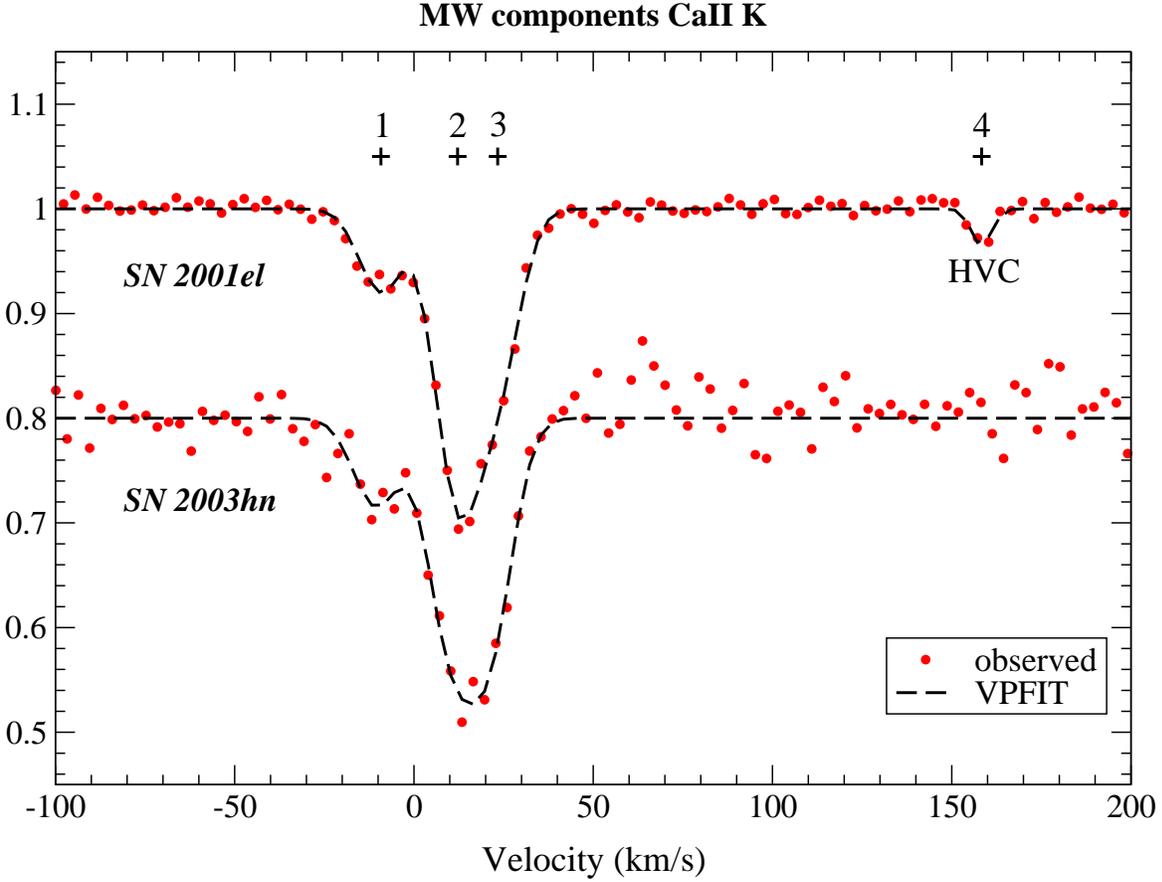}
\end{center}  
\caption{
Interstellar atomic Milky Way components of \ion{Ca}{ii}~K in the line-of-sight
towards SN\,2001el and SN\,2003hn. The black dashed lines refer to the fits 
by VPFIT,
and the grey dots to the observed UVES spectra. For velocities, column
densities and Doppler parameters of the fits we refer to Table~\ref{tb:ISfit}.  
The numbered symbols ($+$) at the top indicate the positions of the fitted velocity
components. The lower spectrum has been displaced vertically for clarity.
A high velocity cloud at $\sim$ 160~km s$^{-1}$ can also be seen towards SN\,2001el (component 4).} 
\label{fig:IS_MW}
\end{figure*}

The Ca\,II H\&K lines are clearly detected also in the MW. On the
sky, the two lines-of-sight given by the two supernovae are separated
by about 69 arcseconds. The MW line profiles also look very similar
with two strong components centered at 
heliocentric radial velocities of about 10 to 20~km~s$^{-1}$ 
(Fig.~\ref{fig:IS_MW}; Table~\ref{tb:ISfit}). Towards SN\,2001el
we also detect a high velocity component at $\sim$158~km~s$^{-1}$
which can not be seen in the noisier data obtained towards SN\,2003hn.

\begin{figure*}[t]
\begin{center}
\includegraphics[width=130mm, angle=-90, clip] {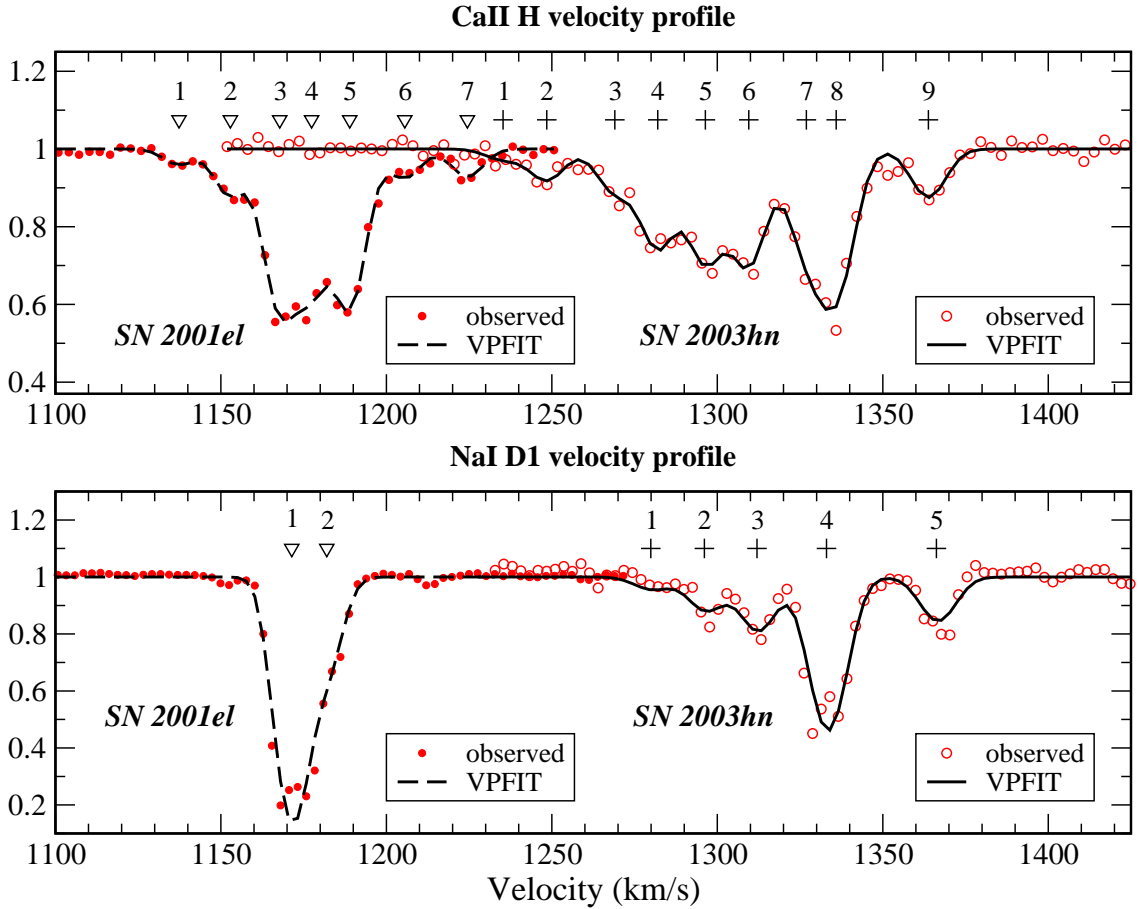}
\end{center}  
\caption{Interstellar atomic lines of \ion{Ca}{ii}~H (top) and 
\ion{Na}{i}~D1 (bottom) 
towards SN\,2001el and
SN\,2003hn. The black lines 
(dashed for SN\,2001el and solid for SN\,2003hn) are 
the fits with VPFIT, and the grey symbols 
(filled dots for SN\,2001el and open circles for SN\,2003hn) 
are the observed UVES spectra. The numbered symbols ($\triangledown$
for SN\,2001el and $+$ for SN\,2003hn) at the top indicate the
positions of the fitted velocity components. 
For velocities, column densities and
Doppler parameters of the fit we refer to  Table~\ref{tb:ISfit}. 
}
\label{fig:IS_NGC}
\end{figure*}

For the NGC\,1448 components of Ca\,II H\&K, the differences 
between the two lines of sight
are apparent. 
Towards SN\,2001el, the strongest components 
are redshifted by $\sim$1170~km~s$^{-1}$, 
while the strongest components towards
SN\,2003hn have a redshift of $\sim$1340~km~s$^{-1}$
(Fig.~\ref{fig:IS_NGC}; Table~\ref{tb:ISfit}). 
This is consistent with the measured
heliocentric velocity of 1168~km~s$^{-1}$, 
with the difference between the two positions in the galaxy reflecting
the rotation velocity ($\sim$193~km~s$^{-1}$; \cite{MF96}). 

To further analyze these line profiles we have used the
program 
VPFIT\footnote{By R. Carswell on www.ast.cam.ac.uk/$\sim$rfc/vpfit.html}
which fits multiple Voigt profiles to multiple line components. We
constrained all the lines (of the same species and ionization state)
to have the same width, where the theoretical line is convolved with
the instrument resolution 
($\sim$6~km s$^{-1}$) 
before doing the fitting. The program 
adds components in an iterative way until an acceptable fit is
found. This initial guess can then be adjusted interactively.
The program uses a least-square fitting method to obtain 
the best fit, and 
in the end VPFIT provides velocities, widths and column densities 
for each line component of the ions. 
The obtained
results are given in Table~\ref{tb:ISfit}, and 
some fits performed by VPFIT 
are shown in Fig.~\ref{fig:IS_NGC}.

\subsection{Diffuse Interstellar Bands}

A very interesting feature of our spectra is the abundance of
extragalactic DIBs. We detect more than a dozen of bands throughout
the spectra of SN\,2001el. 
A list of detected lines, identifications, observed central
wavelengths ($\lambda_{\rm observed}$), velocities (v$_{\rm DIB}$),
equivalent widths (EW) and Full Width Half Maxima (FWHM) is given in
Table~\ref{tb:DIB}.

\begin{figure*}[t]
\begin{center}
\includegraphics[width=130mm, angle=-90, clip] {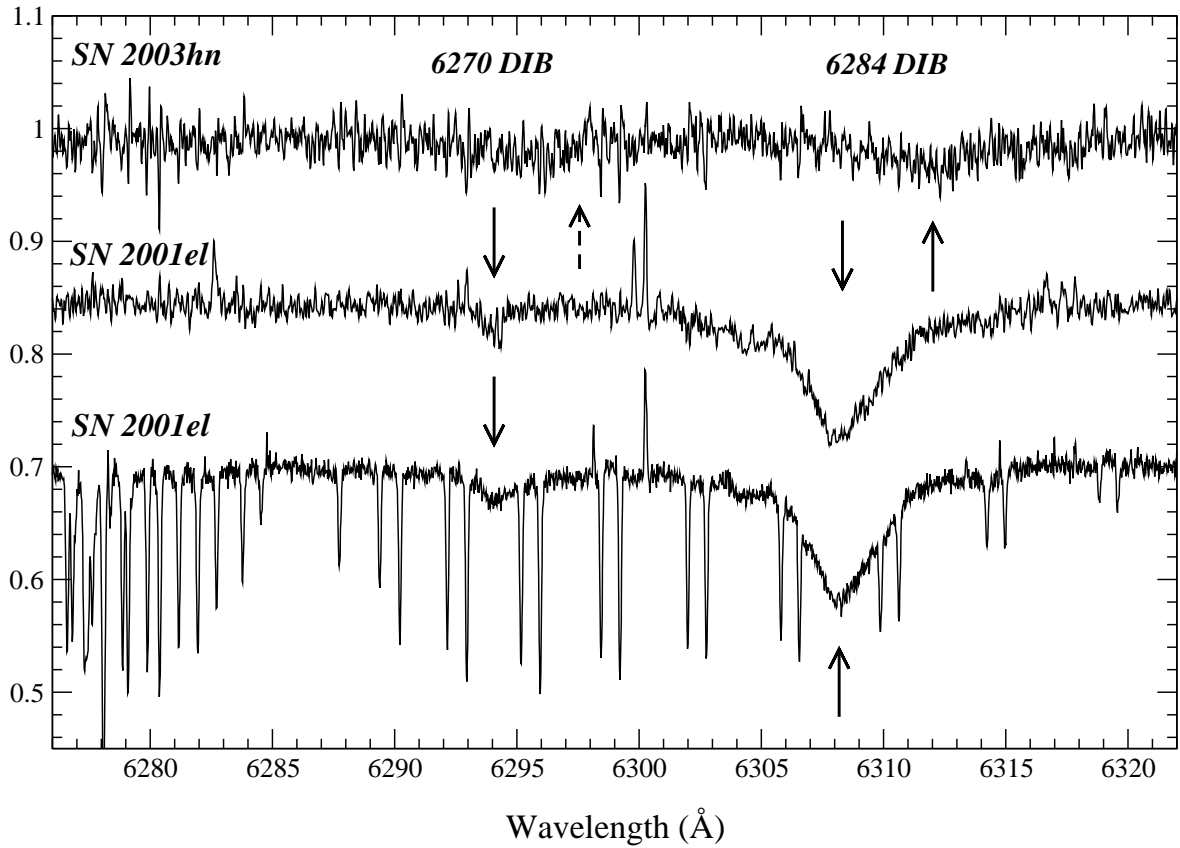}
\end{center}  
\caption{
The $\lambda$\,6284 DIB towards SN\,2001el is redshifted to 6308~\AA~and is clearly separated
from the narrow telluric O$_2$ complex at $\sim$6278~\AA\ (bottom). Note also the $\lambda$\,6270
DIB at 6294~\AA. For illustrative purposes we show both the telluric uncorrected 
and corrected $\lambda$\,6284 DIB for SN\,2001el (bottom and middle). 
These spectra have been displaced vertically for clarity. 
For SN\,2003hn only the telluric corrected spectrum is shown (top).  In this sight line we 
detect the  $\lambda$\,6284 DIB at $\lambda$\,6312.} 
\label{fig:6284}
\end{figure*}

There are many advantages in searching for DIBs in an extragalactic
supernova spectrum. All wavelengths are conveniently redshifted to
avoid confusion with any MW components. 
In our spectra, we detect no DIBs from the MW.
The strong DIB at $\lambda$\,6284 is often blended with a 
telluric O$_{2}$ complex in the MW.
Here the DIB feature is redshifted to 6308~\AA , and the high
resolution clearly separates the narrow telluric lines
(Fig.~\ref{fig:6284}). 
There is also no
contamination from intrinsic narrow lines that needs to be modeled
in the supernova spectrum, as opposed to using stars as background sources.
However, the supernova spectrum is made up of a superposition of
numerous broad lines. This is well suited as a quasi-continuum against
which to detect narrow DIBs, but very broad DIBs are not so easy
to disentangle. We have therefore not been able to clearly identify
DIBs with FWHM broader than $\gtrsim$10~\AA. For example, the usually
very strong DIB at $\lambda$\,4428  can not be securely identified.
We emphasize that this is not to be interpreted as evidence for
absence of such broad DIBs (see e.g., \cite{E97}).  

The sample of lines listed in Table~\ref{tb:DIB} were searched among
the DIBs tabulated by Herbig (1995).  From this table, we have
searched and detected all the 
strong (EW$>$200~m\AA ) lines with
FWHM$<$7~\AA~between 4000 and 8000~\AA.  
The line at $\lambda$\,7724 only became obvious
after division with a standard star to cancel out the
telluric lines.  
All these 9 lines have a central depth (A$_{c}$) $\gtrsim$0.07, as defined
by Herbig (1995). 
We therefore searched also for all the other
tabulated  DIBs that meet this criterion.

\begin{figure*}[t]
\begin{center}
\includegraphics[width=130mm, angle=-90, clip] {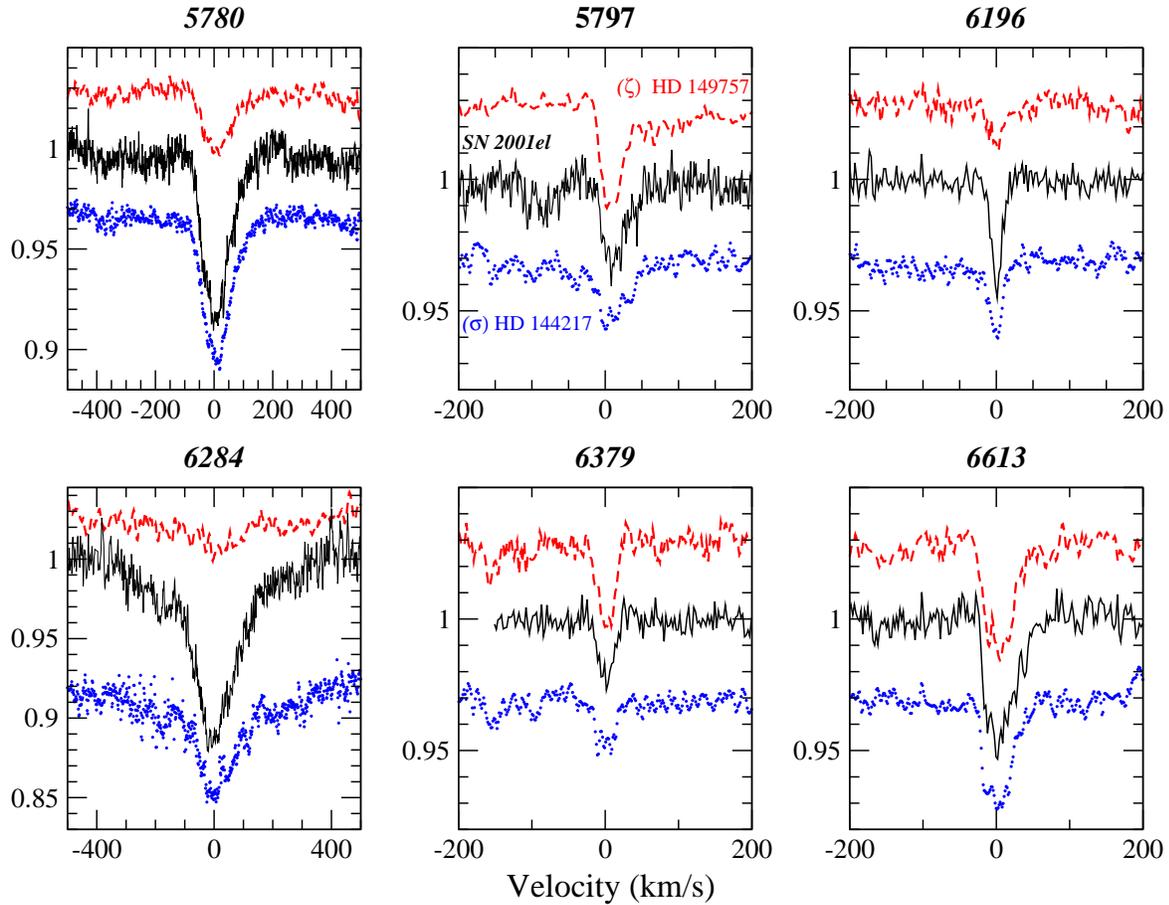}
\end{center}  
\caption{The six panels show, on a velocity scale centered on the DIBs, 
some of the most important and well studied DIBs 
($\lambda\lambda$\, 5780, 5797, 6196, 6284, 6379 and $\lambda$\,6613)  
as observed towards SN\,2001el. 
Overplotted are the observed FEROS DIB spectra towards the 
$\zeta$ type cloud HD\,149757 [top, E(B-V)=0.32 mag] and the $\sigma$ type
cloud HD\,144217 [bottom, E(B-V)=0.20 mag].
These spectra are not scaled, but 
shifted vertically for clarity.
For measured values of the DIB velocities, equivalent widths and
full-width half maxima see Table~\ref{tb:DIB}.} 
\label{fig:DIB}
\end{figure*}

Apart from the broad, shallow $\lambda$\,4428 feature, as discussed above, we
detect also the other 4 DIBs ($\lambda\lambda$ 6196, 6379, 6661 and 6993)  
with A$_{c}$$\gtrsim$0.07 in the  wavelength range given
above. This includes the narrow line at  $\lambda$\,6196,
which is clearly detected.  After applying the telluric correction we
also detect the weak $\lambda$\,6661 and $\lambda$\,6993 DIBs.
Six conspicious DIBs towards SN\,2001el are displayed in Fig.~\ref{fig:DIB}.
Longwards of 8000 \AA , there are 3 potentially strong DIBs
($\lambda\lambda$\ 8621, 9577 and 9632) according to the list of
Herbig (1995), but we were unable to detect any of these lines.  This
region was only observed during our first epoch of observations.
In this study 
we will use the clear detections to compare our
observations with DIB observations in the MW and in other galaxies.

The spectra of SN\,2003hn do not display the same multitude of
DIBs as the line of sight towards SN\,2001el. 
We were able to detect only 2 DIBs, 
the $\lambda$\,5780 and $\lambda$\,6284
with any confidence
(see Table~\ref{tb:DIB} and Fig.~\ref{fig:6284}).

\subsection{Extinction}
\label{sec:reddening}

There are many ways to estimate the amount of extinction towards an 
astronomical object. In supernova research, estimates are often made from 
the equivalent widths of the interstellar 
Na~I~D lines - even from low resolution 
spectroscopy (e.g., Turatto et al. 2003 and references therein). 
In this work we have high quality high resolution spectra and are 
able to deduce the actual column densities for these lines. Moreover, 
for Type Ia supernovae, an estimate of the reddening can be directly obtained 
from the supernova colors. 
Therefore, this dataset allows
a comparison between the different methods. 

\subsubsection{Equivalent Width and Column densities}

The use of the Na I D EW 
to estimate the amount of reddening 
(e.g., \cite{Barb90}; \cite{Turatto03})
often assumes that the
effects of saturation are negligible. 
However, this is not valid for our observations. Table~\ref{tb:EWs} shows 
our measurements of the EW and column densities 
towards the two SNe. 

In this table we have first assumed an optically thin line for which the
column density is directly proportional to the equivalent width
(e.g., \cite{Spitzer78}).  
In the lower part of this table 
we also show the results from the so called doublet ratio (DR) technique
(e.g., \cite{Somerville88}), as well as the total column densities
derived with VPFIT.
It is clear that the optically thin
approximation is not valid for the saturated doublet lines
towards NGC\,1448. 
Taking the saturation into account via either the
DR method or the component fitting procedure (VPFIT) gives
column densities that are mutually consistent. 
A simple Na I D EW 
approach to estimate the amount of reddening towards a supernova can 
thus
give substantial errors (see also discussions by e.g., \cite{MZ97}, 
\cite{Fassia00}~and \cite{Smartt02}).
We will use the 
column densities derived by VPFIT
to estimate the amount of reddening below.

\subsubsection{Estimate of $E(B-V)$}

Adopting the relation from Hobbs (1974) we can use our measured sodium column 
densities to directly derive the reddenings towards the supernovae. 
This gives $E(B-V)_{\rm host}=0.15$ and 0.12 mags 
for SN\,2001el and SN\,2003hn,
respectively. The values for the MW components are 
$E(B-V)_{\rm MW}=0.021$ and 0.017 mags, respectively.
For the MW components we can directly compare this to the value derived by
Schlegel et al. (1998), $E(B-V)_{\rm MW}=0.014$ mag.

Alternatively (following e.g., \cite{Fassia00}), 
one can convert the sodium column density 
to hydrogen column density (\cite{Ferlet85}) 
assuming a MW gas-to-dust ratio, 
and then derive the color excess (\cite{Bohlin78}). 
This gives  $E(B-V)_{\rm host}=0.18$ mag
towards SN\,2001el.

Another way to measure the reddening towards SN\,2001el is from the
supernova light curve itself. It has been established that Type Ia
supernovae display a uniform intrinsic color evolution from 30 to 90
days past maximum (\cite{Lira95}; \cite{Phillips99}).
The light curve of SN\,2001el has been
very well monitored by 
Krisciunas et al. (2003).
They obtained various
estimates of the color excess using the intrinsic color of the supernova. 
Using the light curve tail, a value of
$E(B-V)_{\rm total}=0.253\pm0.063$ mag is reported, whereas
an average of all the six different methods used by 
Krisciunas et al. 
gives 
$E(B-V)_{\rm total}=0.185\pm0.07$ mag. 
Here the error is 
a combination of
the propagated errors of the different estimates 
and the standard deviation in the estimates themselves. 
When we compare this with the total reddening towards SN\,2001el derived 
from the sodium lines, we find a good agreement within the 
errors.
Below we will use $E(B-V)_{\rm host}=0.18\pm0.08$ mag
for SN\,2001el. This value encapsulates most of the estimates given 
in this section.

\section{Discussion}

\subsection{Line profiles}
In Fig.~\ref{fig:DIB} we show the line profiles of several conspicuous
DIBs in NGC\,1448. 
The spectra
are shown in velocity scale in kilometers  per second
with respect to the central wavelength indicated in the figure. 

The $\lambda$\,6613 line exhibits a much steeper blue side of the
profile, and  this can also be perceived, for example, in the
$\lambda$\,5797 line.   The $\lambda\lambda$\,6284 and 5780 DIBs have
quite similar line profiles, and are clearly not simple
Gaussians. 
The observed profiles actually show a strong resemblance to those seen 
towards galactic single cloud lines of sight. To illustrate this we also 
show  two typical galactic lines of sight that represent the so called 
$\sigma$ (HD\,144217) and $\zeta$ (HD\,149757) type diffuse clouds
(see e.g., \cite{krelo95}). 
These galactic spectra were obtained with the FEROS instrument by one of us 
(LK). These lines of sight have $E(B-V)$ of 0.20 mag and 0.32 mag, 
respectively. 
Note that the SN\,2001el DIBs show the same asymmetries in the profiles as 
the MW DIBs.

\subsection{DIB velocities}

\begin{figure*}[t]
\begin{center}
\includegraphics[width=140mm, angle=-90, clip] {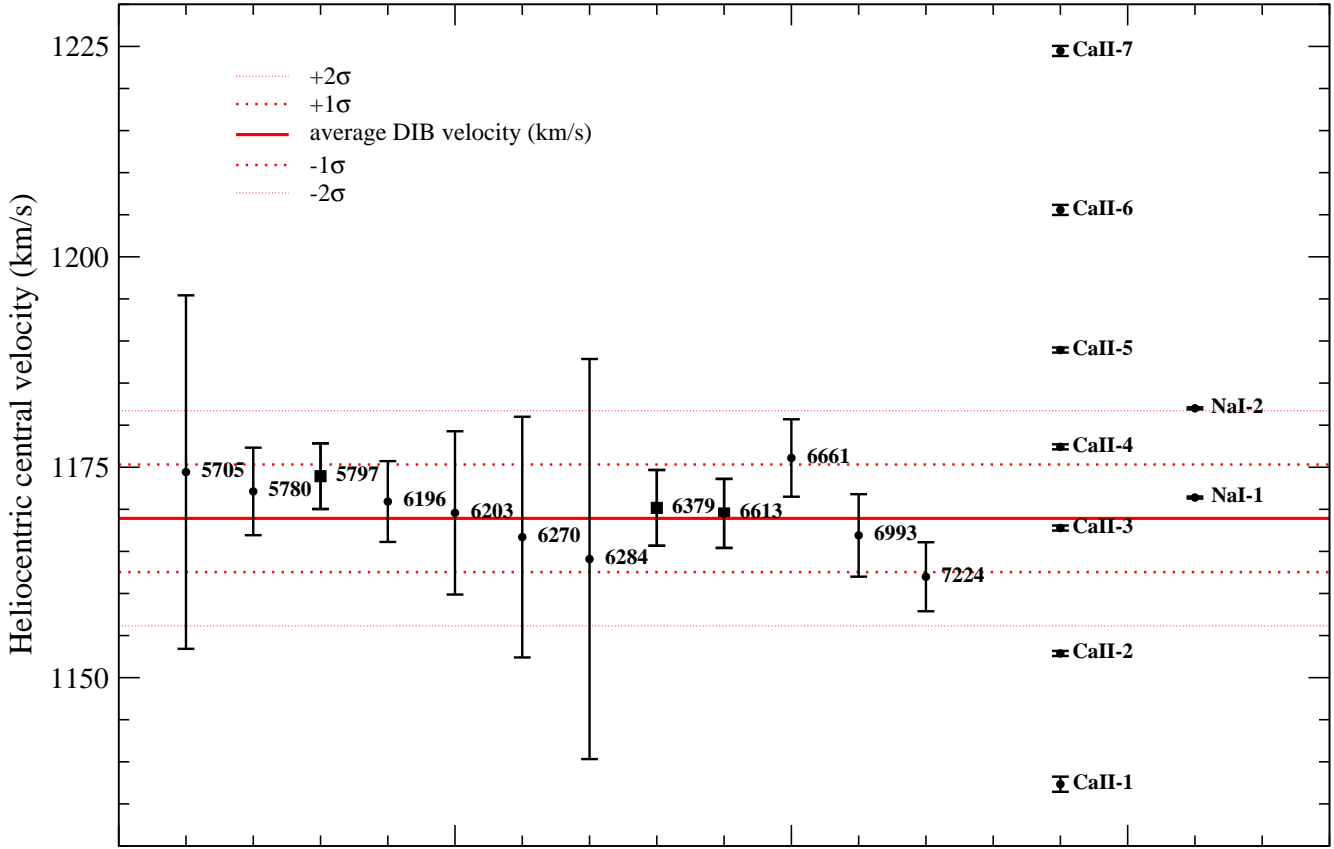}
\end{center}  
\caption{The central heliocentric velocities of the 12 DIBs observed
towards SN\,2001el.  The square symbols indicate the $\lambda$\,5797,
$\lambda$\,6379 and $\lambda$\,6613 family (\cite{C97}).
The derived velocities of the individual Ca\,II and Na\,I components are
also indicated. 
The
central velocities for the strong, narrow DIBs are well defined,
whereas those of the broader and/or weaker are less stringent.
The average DIB velocity coincides with the strongest interstellar
line components Ca\,II-3 and Na\,I-1, and within 2$\sigma$ also with
the components Ca\,II-4 and Na\,I-2.  Although these two component-pairs
have very similar column densities (and corresponding
N(Na\,I)/N(Ca\,II) ratios),  one velocity component seems to be favoured by
the DIBs. 
The SN\,2001el DIBs (Table~4) show no broadening 
with respect to the single cloud lines of sight (Fig.~\ref{fig:DIB}), 
and might thus be expected to originate within a small velocity range.
}
\label{fig:velocities}
\end{figure*}

In Fig.~\ref{fig:velocities} we explore the link between DIB velocities and
atomic line  velocities. It is clear that the DIB carriers (of the
narrow DIBs) are, in velocity, closely related to the strongest Ca\,II
and Na\,I line components. The four narrowest DIBs
($\lambda\lambda$\,6379, 6613, 6661 and 6993) have well determined
central velocities that can be assigned, within the errors, to
components 3 and 1 of the Ca\,II and Na\,I profiles, respectively. 
This velocity coincidence indicates that the carriers of these DIBs are 
physically associated, and probably located within the same cloud in NGC 1448.

Since we also observe no broadening of the profiles with 
respect to the single cloud DIBs towards  HD\,144217 and HD\,149757 
(Fig.~\ref{fig:DIB})
we conclude that the DIBs towards SN\,2001el 
primarily form in  a single gas-rich layer, 
indicated by these strong absorption 
components of ionized Ca and neutral Na.

\subsection{DIB ratios}	

Two potentially 
important DIBs for the determination of the ionization balance are the
ones at $\lambda$\,5797 and $\lambda$\,5780.  
According to Cami et al. (1997)
the $\lambda$\,5780 DIB has a
higher ionization potential than the $\lambda$\,5797 DIB, and thus
reaches its maximum only with a stronger UV field.
For SN\,2001el the $\lambda$\,5780 DIB is very strong compared to the
$\lambda$\,5797 DIB. This behavior is 
indicative of a
so called $\sigma$ type cloud like environment
(Fig.~\ref{fig:DIB}). 
In such a cloud Ca\,I and simple interstellar molecules 
(CH, CN) are very weak or undetectable. 
This is also true for our observations, 
where the $3\sigma$ upper limits on CH, CN are
7 and 10 m\AA~, respectively.

From Table~\ref{tb:DIB} we can
compare the 5797/5780 ratio for different galactic and
extragalactic targets.
The denser single cloud towards 
HD\,149757 has a relatively high ratio 
($\sim0.4$), while 
for NGC\,1448,  the LMC and HD\,144217
we see ratios of about 0.15, which
within the interpretation of Cami et al. (1997),
indicate a somewhat higher UV field.

\subsection{Extragalactic DIBs}

As mentioned in the introduction, 
extragalactic DIBs have only been observed in a few cases.
In this study the quality of the data allows a detailed comparison with 
the DIBs in the Milky Way.

For SN\,2001el we 
compare all the DIBs with an  'average cloud' in the MW 
as given in Table~4.
Although the DIBs against SN\,2001el appear relatively strong, 
this could just be due to the 
uncertainty in the determination of the reddening. The DIB strengths would
be similar to those of the average cloud for an $E(B-V)\sim0.3$ mag, 
which is still within the error budget.

In fact, as illustrated in Fig.~\ref{fig:DIB} and given in Table~4, we find 
that the properties of the DIBs in NGC\,1448 closely mimic those observed 
towards HD\,144217. Both the line profiles and the relative strengths are
similar for these lines of sight, illustrating the potential for 
studying how DIB carriers behave in different extragalactic environments.
It could also indicate that very similar local environmental conditions 
pretain in those different lines of sight. 

\begin{figure*}[t]
\begin{center}
\includegraphics[width=130mm, angle=-90, clip]{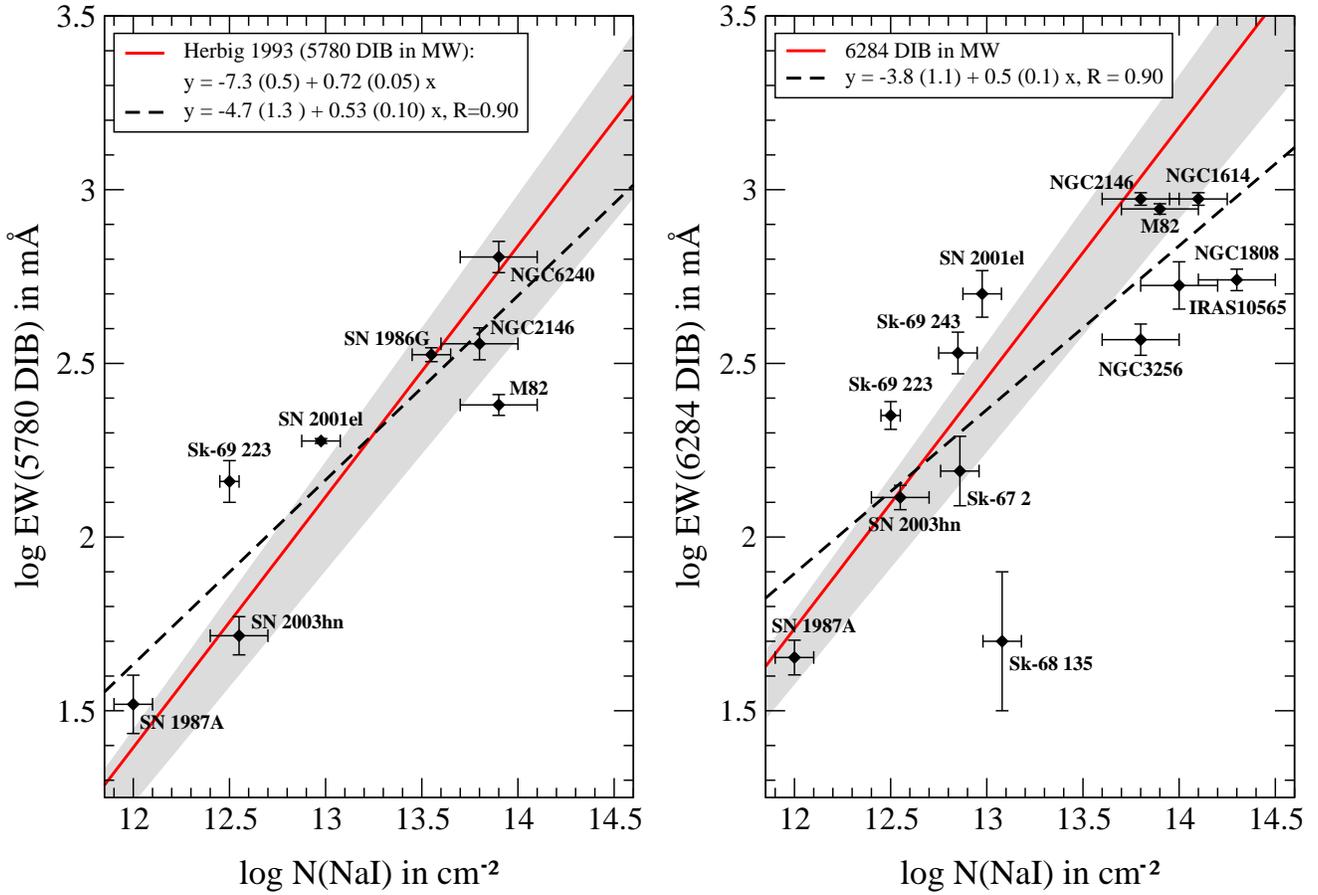}
\end{center}  
\caption{The equivalent widths of the extragalactic 
DIBs $\lambda$ 5780 ({\it left panel}) and $\lambda$ 6284 ({\it right panel}) 
are plotted versus the extragalactic Na\,I column
densities in their line-of-sights. Total column densities have been derived
from the 
Na\,I line and do not take into account individual
components. SN\,1987A data are from Vladilo et al. (1987) and Vidal-Majar et
al. (1987).  Sk-69 223 (LMC) data are from Cox et al. (in preparation), and Sk-67 2, Sk-68 135 from 
Ehrenfreund et al. (2002). 
SN\,1986G data are from D'Odorico et al. (1989).  
The remaining extragalactic 
points are starburst galaxies from Heckman \& Lehnert (2000).  
The solid line in the left panel is the relationship from Herbig (1993) for the MW.
The grey region illustrates the 
1$\sigma$ uncertainty region for that relation.
In the right panel, this relationship has been converted for the 
$\lambda$ 6284 DIB assuming $<5780/6284>=2.2$, 
as done in Heckman \& Lehnert (2000).
The dashed lines are the best linear fits to the extragalactic data.
 }
\label{fig:EWvsNNaI}
\end{figure*}

We have also compiled a sample of extragalactic DIBs to compare
with the properties of the MW DIBs. 
In Fig.~\ref{fig:EWvsNNaI} we extend the exercise 
of Heckman \& Lehnert (2000) and plot
the EW of the $\lambda\lambda$\,5780 and 6284 DIBs 
versus the total sodium column density.
The extra galactic sight lines 
show, for their respective reddenings, 
similar EWs compared to the galactic average.  
In the left panel we have not included the detections 
classified as tentative by Heckman \& Lehnert (2000). 
These would appear significantly below the fitted line.

Herbig (1995) summarized that extragalactic DIBs did not show conclusive 
evidence for any variation of DIB strengths versus color excess, partly due
to the large scatter in the Galactic data. This seems to hold also for the
data presented here.

\subsection{Location of the absorbing gas}

The information we have gathered could potentially 
provide some clues on the location and 
origin of the material in which the DIBs are produced. It may even 
be of interest to investigate to which extent this material is physically 
connected to the local supernova environment.

Jenkins et al. (1984) noted for SN\,1983N 
that the presence of neutral sodium and singly ionized calcium 
argue 
against absorbing gas close to the supernova location. 
Since towards SN\,2001el these ISM lines are correlated with the DIBs 
(Fig.~\ref{fig:velocities}) the same argument would imply 
that the DIBs are not directly located in the supernova environment.
Also, we measure no variability in the DIBs between the two epochs
(e.g., the EW for the $\lambda$ 6613 DIB is stable to about 6\%).
This does not favor a scenario were the DIBs are produced in gas
closely associated with the supernova itself, and is 
consistent with conclusions from supernovae Type Ia
investigations arguing that the dust dimming the supernovae is
generally interstellar rather than circumstellar (e.g., \cite{RPK96}).

The ToO programme behind these data has also observed a few other 
supernovae of various types; SNe 2000cx (Type Ia), 
2001ig (IIb), 2003bg (II).
None of these showed any signatures ($>2\sigma$) of the strongest 
DIBs ($\lambda\lambda$ 5780, 6284). SNe 2000cx and 2001ig would have 
revealed bands similar in strength
to those seen towards SN\,2003hn. 
In the noisier spectrum of SN 2003bg we would only 
have detected ($\sim2\sigma$) bands as strong as those seen towards 
SN\,2001el.

For the other supernovae where 
DIBs have been seen (SNe 1986G, 1987A, 1989M) there is also no
clear correlation between the DIB strengths and the supernova type.

To truly compare the different supernova sight lines 
would require a more thorough investigation of the
host galaxies, including for example their metallicities. This is 
beyond the scope of this investigation - where the main aim was instead to
demonstrate the potential of probing DIBs in external galaxies using
supernovae as transitory luminous probes for high resolution spectroscopy.

\section{Conclusions}

Using high resolution spectroscopy of two emerging supernovae in NGC 1448 
we have detected a number of extragalactic atomic interstellar lines. 
Towards SN\,2001el we also detected a large number of Diffuse Interstellar 
Bands.

We have compared the properties of the DIBs in NGC 1448 with those of
the DIBs observed both in our own Galaxy and in other galaxies.
These observations probe the most distant system where a larger number of DIBs
has been analyzed in such a detail. These DIBs show many similarities with 
DIBs within the Milky Way, 
especially with those seen towards the $\sigma$-type cloud HD\,144217.
This shows the potential for modern telescopes to investigate how
DIB carriers follow common chemical and 
physical pathways throughout the universe.

We have shown that the DIBs towards SN\,2001el are associated in velocity 
space with specific components of the atomic interstellar lines. 
We observe no DIB strength time variability on time scales shorter than a
week, nor do we see any direct connection between DIB properties and
supernova type. 

We have also probed the extinction towards the supernovae in several
different ways. 
Taking the saturation of the interstellar sodium lines 
into account in our high-resolution data gives a reddening
estimate consistent with color excess measurements from the Type Ia 
SN\,2001el itself.\\

{\em Acknowledgements.}
These observations were obtained in ToO service mode at the VLT. 
We wish to thank the Paranal 
staff for all the help.
We also thank J. Fynbo for comments on the manuscript. 
C. Fransson, E. Baron and K. Nomoto were helpful in writing the 
original UVES proposals. NC acknowledges NOVA for financial support and
SM acknowledges financial support from the ``Physics of Type Ia SNe'' RTN.

\clearpage

\begin{table}
\caption{Log of VLT/UVES observations of SN\,2001el.}
\begin{tabular}{lllllll}
\hline
\hline
Date    & MJD        & Exp.  & Airmass & Seeing$^a$ & Set-up & Slit width\\ 
(01 09) & (52000+)   & (s)   &      &  (arcsec)  &        & (arcsec) \\
\hline

21 & 173.22 & 2400 & 1.30 & 0.82 & 390+564$^b$ & 0.8 \\
21 & 173.25 & 2400 & 1.19 & 0.79 & 390+564 & 0.8 \\
21 & 173.28 & 2400 & 1.12 & 0.65 & 437+860$^c$ & 0.8  \\
21 & 173.31 & 2400 & 1.08 & 0.85 & 437+860 & 0.8 \\

26 & 178.36 & 1200 & 1.10 & 1.43 & 346+580$^d$ & 0.7 \\
26 & 178.39 & 1200 & 1.12 & 1.43 & 346+580 & 0.7 \\

28 & 180.22  & 3000 & 1.20 & 1.05 & 390+564 & 0.8 \\
28 & 180.26  & 3000 & 1.11 & 1.08 & 390+564 & 0.8 \\
28 & 180.29  & 3000 & 1.07 & 1.12 & 390+564 & 0.8 \\

\hline
\end{tabular} \\
\begin{tabular}{lll}
$^a$ \ Seeing from the DIMM-monitor. \\
$^b$ \ Setting 390+564 covers wavelength ranges 3260-4450, 4580-6680~\AA. \\
$^c$ \ 437+860 covers wavelength ranges 3730-4990, 6600-10600. \\
$^d$ \ 346+580 covers wavelength ranges 3030-3880, 4760-6840. \\
\end{tabular}
\label{t:obs01el}
\end{table}

\begin{table}
\caption{Log of VLT/UVES observations of SN\,2003hn.}
\begin{tabular}{lllllll}
\hline
\hline
Date    & MJD       & Exp.  & Airmass & Seeing$^a$ & Set-up & Slit width\\
(03 08) & (52000+)  & (s)   &      & (arcsec)   &        & (arcsec)  \\
\hline

31 & 882.26 & 1400 & 1.42 & 0.57 & 390+564$^b$ & 0.8 \\
31 & 882.28 & 1400 & 1.32 & 0.61 & 390+564 & 0.8 \\
31 & 882.30 & 1400 & 1.25 & 0.52 & 390+564 & 0.8 \\
31 & 882.33 & 1400 & 1.15 & 0.51 & 437+860$^c$ & 0.8  \\
31 & 882.34 & 1400 & 1.11 & 0.55 &  437+860 & 0.8 \\
31 & 882.36 & 1400 & 1.09 & 0.53 &  437+860 & 0.8 \\

\hline
\end{tabular} \\
\begin{tabular}{lll}
$^a$ \ Seeing from the DIMM-monitor. \\
$^b$ \ Setting 390+564 covers wavelength ranges 3260-4450, 4580-6680~\AA. \\
$^c$ \ 437+860 covers wavelength ranges 3730-4990, 6600-10600. \\
\\
\end{tabular}
\label{t:obs03hn}
\end{table}

\clearpage

\begin{table*} {\footnotesize
\caption{
VPFIT parameters (velocity $v$, Doppler parameter $b$ and column density $N$) for the components of the observed atomic interstellar lines of \ion{Ca}{ii}, \ion{Ti}{ii}, \ion{Na}{i} and \ion{K}{i} towards SN\,2001el and SN\,2003hn. In the last column we give the total column density, summed over the individual components.
}
\begin{center}
\begin{tabular}{lcccccc}
\hline
\hline
&Line$^a$   & Component 	& $v$			& $b$$^b$ 	 	& log $N$ 			& total log $N$        \\
&	  	&	     		 & (km s$^{-1}$)	& (km s$^{-1}$)		& (cm$^{-2}$)  	& (cm$^{-2}$)       \\ \hline
\multicolumn{3}{l}{\bf MW components:} & & & \\ 
SN\,2001el&\ion{Ca}{ii}& 1 & 23.3	$\pm$ 0.3  &  7.6 $\pm$ 0.3 & 11.57 $\pm$ 0.03   &  \multirow{3}*{\resizebox{5mm}{8mm}{\Large \} } } \multirow{3}*[0mm]{12.12 $\pm$ 0.03} \\  
&\ion{Ca}{ii}    &  2       & 12.2	$\pm$ 0.3  &		       & 11.86 $\pm$ 0.02  &        \\  
&\ion{Ca}{ii}    &  3       & -9.2	$\pm$ 0.6  &		       & 11.28 $\pm$ 0.02    &        \\  
&\ion{Ca}{ii}    &  4       & 158.4  $\pm$ 1.2  &  1.7 $\pm$ 4.1 & 10.60 $\pm$ 0.08 &  \multirow{1}*{\resizebox{5mm}{2mm}{\Large \} } } \multirow{1}*[0mm]{10.60 $\pm$ 0.08}  \\ 
&\ion{Ti}{ii}    &  1       & 11.2	$\pm$ 12.0 &  9.7 $\pm$ 7.3 & 11.59 $\pm$ 0.63   &  \multirow{2}*{\resizebox{5mm}{5mm}{\Large \} } } \multirow{2}*[0mm]{11.83 $\pm$ 0.50}  \\  
&\ion{Ti}{ii}    &  2       & 21.4	$\pm$ 8.1  &		       & 11.45 $\pm$ 0.82   &        \\  
&\ion{Na}{i}     &  1       & 17.2	$\pm$ 0.6  & 10.1 $\pm$ 1.0 & 11.21 $\pm$ 0.03 &   \multirow{1}*{\resizebox{5mm}{2mm}{\Large \} } } \multirow{1}*[0mm]{11.21 $\pm$ 0.03} \\ 
SN\,2003hn&\ion{Ca}{ii}    &  1       & 20.74 $\pm$ 1.2  &  8.0 $\pm$ 1.3 & 11.75 $\pm$ 0.09 	 &  \multirow{3}*{\resizebox{5mm}{8mm}{\Large \} } } \multirow{3}*[0mm]{12.12 $\pm$ 0.06} \\
&\ion{Ca}{ii}    &  2       & 10.0	$\pm$ 1.8  &		       & 11.73 $\pm$ 0.09   &        \\
&\ion{Ca}{ii}    &  3       & -10.7  $\pm$ 1.8  &		       & 11.32 $\pm$ 0.07      &        \\
&\ion{Na}{i}  &  1       & 19.3	$\pm$ 1.2  &  5.9 $\pm$ 1.9 & 11.04 $\pm$ 0.08 &   \multirow{1}*{\resizebox{5mm}{2mm}{\Large \} } } \multirow{1}*[0mm]{11.04 $\pm$ 0.08} \\ 
\multicolumn{7}{l}{}\\
\multicolumn{3}{l}{\bf NGC\,1448 components:} & & & \\
SN\,2001el&\ion{Ca}{ii} &1 & 1137.3 $\pm$ 0.9  & 5.3 $\pm$ 0.2 & 11.08 $\pm$ 0.06       &  \multirow{7}*{\resizebox{5mm}{20mm}{\Large \} } } \multirow{7}*[0mm]{12.79 $\pm$ 0.01} \\
&\ion{Ca}{ii}    &  2       & 1152.9 $\pm$ 0.3  &		      & 11.64 $\pm$ 0.02   &   \\
&\ion{Ca}{ii}    &  3       & 1167.8 $\pm$ 0.3  &		      & 12.26 $\pm$ 0.02      &        \\
&\ion{Ca}{ii}    &  4       & 1177.4 $\pm$ 0.3  &		      & 12.17 $\pm$ 0.02      &        \\
&\ion{Ca}{ii}    &  5       & 1188.9 $\pm$ 0.3  &		      & 12.26 $\pm$ 0.01   &        \\
&\ion{Ca}{ii}    &  6       & 1205.6 $\pm$ 0.6  &		      & 11.38 $\pm$ 0.03   &        \\
&\ion{Ca}{ii}    &  7       & 1224.5 $\pm$ 0.6  & 		      & 11.33 $\pm$ 0.03   &        \\
&\ion{Na}{i}  &  1       & 1171.4 $\pm$ 0.1  & 5.0 $\pm$ 0.2 & 12.68 $\pm$ 0.03  &  \multirow{2}*{\resizebox{5mm}{5mm}{\Large \} } } \multirow{2}*[0mm]{12.76 $\pm$ 0.03} \\
&\ion{Na}{i}  &  2      & 1182.0 $\pm$ 0.1  &		      & 11.96 $\pm$ 0.03    &        \\
&\ion{Ti}{ii}    &  1      & 1148.0 $\pm$ 3.6  & 12.9 $\pm$ 3.4 & 11.78 $\pm$ 0.11   & \multirow{4}*{\resizebox{5mm}{10mm}{\Large \} } } \multirow{4}*[0mm]{12.69 $\pm$ 1.40} \\  
&\ion{Ti}{ii}    &  2       & 1174.7 $\pm$ 25.5 &		       & 12.40 $\pm$ 1.98  &        \\  
&\ion{Ti}{ii}    &  3       & 1181.6 $\pm$ 18.6 &		       & 12.15 $\pm$ 3.42  &        \\  
&\ion{Ti}{ii}    &  4       & 1203.1 $\pm$ 6.9  &		       & 11.64 $\pm$ 0.32   &        \\	    
&\ion{K}{i}	   &  1       & 1166.6 $\pm$ 0.6  & 1.4 $\pm$ 0.9 & 11.02 $\pm$ 0.07    & \multirow{2}*{\resizebox{5mm}{5mm}{\Large \} } } \multirow{2}*[0mm]{11.29 $\pm$ 0.05} \\  
&\ion{K}{i}	   &  2       & 1175.6 $\pm$ 0.6  &		      & 10.96 $\pm$ 0.06    &        \\  
\hline			

SN\,2003hn:&\ion{Ca}{ii}& 1 & 1235.3 $\pm$ 3.6  &  6.9 $\pm$ 0.5    & 11.08  $\pm$ 0.16  & 
									\multirow{9}*{\resizebox{5mm}{26mm}{\Large \} } } \multirow{9}*[0mm]{12.89 $\pm$ 0.03} \\
&\ion{Ca}{ii}    &  2       & 1248.5 $\pm$ 1.5  &		       & 11.42  $\pm$ 0.08	 &	  \\
&\ion{Ca}{ii}    &  3       & 1269.0 $\pm$ 1.5  &		       & 11.70  $\pm$ 0.07	 &	  \\
&\ion{Ca}{ii}    &  4       & 1282.0 $\pm$ 0.9  &		       & 12.03  $\pm$ 0.04	&	 \\
&\ion{Ca}{ii}    &  5       & 1296.4 $\pm$ 0.9  &		       & 12.14  $\pm$ 0.03	&	 \\
&\ion{Ca}{ii}    &  6       & 1309.5 $\pm$ 0.6  &		       & 12.14  $\pm$ 0.03	&	 \\
&\ion{Ca}{ii}    &  7       & 1326.9 $\pm$ 1.5  &		       & 12.17  $\pm$ 0.11	&	 \\
&\ion{Ca}{ii}    &  8       & 1335.9 $\pm$ 0.9  &		       & 12.06  $\pm$ 0.12	 &	  \\
&\ion{Ca}{ii}    &  9       & 1363.8 $\pm$ 0.6  &		       & 11.64  $\pm$ 0.04	&	 \\ 
&\ion{Na}{i}  &  1       & 1279.9 $\pm$ 1.5  & 6.4 $\pm$ 0.2	& 11.08 $\pm$  0.08  & \multirow{5}*{\resizebox{5mm}{14mm}{\Large \} } } \multirow{5}*[0mm]{12.55 $\pm$ 0.01}\\ 
&\ion{Na}{i}  &  2       & 1296.1 $\pm$ 0.6  &			& 11.53 $\pm$  0.03  &        \\ 
&\ion{Na}{i}  &  3       & 1312.0 $\pm$ 0.3  &			& 11.75 $\pm$  0.02  &        \\ 
&\ion{Na}{i}  &  4       & 1333.0 $\pm$ 0.3  & 			& 12.32 $\pm$  0.01  &        \\ 
&\ion{Na}{i}  &  5       & 1366.2 $\pm$ 0.3  &			& 11.65 $\pm$  0.02  &        \\ 
&\ion{Ti}{ii}    &  1       & 1281.7 $\pm$ 0.9  & 6.8 $\pm$ 0.8	& 12.17 $\pm$ 0.06 &  \multirow{4}*{\resizebox{5mm}{10mm}{\Large \} } } \multirow{4}*[0mm]{12.64 $\pm$ 0.04} \\
&\ion{Ti}{ii}    &  2       & 1293.4 $\pm$ 1.5  &			& 11.84 $\pm$ 0.11 &        \\
&\ion{Ti}{ii}    &  3       & 1324.6 $\pm$ 1.2  & 			& 12.01 $\pm$ 0.07 &        \\
&\ion{Ti}{ii}    &  4	      & 1337.5 $\pm$ 0.9  &		       & 12.07 $\pm$ 0.06  &        \\
\hline			
\end{tabular} \\
\begin{tabular}{lp{16cm}}
$^a$ & We use the doublet lines of \ion{Ca}{ii}, \ion{Na}{i} and \ion{K}{i} to constrain the least square fitting routine employed by VPFIT.\\
$^b$ & The doppler parameter, $b$, has for each atomic line been set equal for all components. \\
\end{tabular}
\end{center}
\label{tb:ISfit}
}
\end{table*}

\clearpage

\begin{table*}
\caption{DIBs in NGC\,1448 towards SNe\,2001el and 2003hn. 
Central velocities were derived by fitting high resolution 
DIB profiles of single cloud galactic lines of sight to the observed 
DIBs profiles. 
}
\begin{tabular}{l|llll|l|ll|ll}
\hline
\hline
 DIB & \multicolumn{4}{|c|}{NGC1448}& Average $^c$ & HD\,144217$^d$ & HD\,149757$^d$ &LMC $^e$&SN\,1986G $^f$\\ \hline
$\lambda_{\rm rest}$ $^a$ & $\lambda_{\rm observed}$ & v$_{\rm DIB}$  & EW & FWHM & EW$_{\rm scaled}$ & EW & EW & EW  & EW\\
 (\AA)  		 &  (\AA) 	   	    & (km~s$^{-1}$)  & (m\AA) & (\AA) & (m\AA) & (m\AA)  & (m\AA) & (m\AA) & (m\AA) \\
\hline
SN\,2001el: &&&&&&&&&\\
\hline
5705.20   &   5727.55	& 1174.4   $\pm$ 21.0   &  37 $\pm$ 5	 & 2.23 &  17  &  93	   &  --   &  20 $\pm$ 9  &  79 $\pm$ 5\\	  
5780.37   &   5802.97	& 1172.1   $\pm$ 5.2	& 189 $\pm$ 3	 & 2.04 &  104  &  160	   &  66   & 145 $\pm$ 21 & 335 $\pm$ 5\\
5796.97   &   5819.67	& 1173.9  $\pm$ 3.9	&  26 $\pm$ 2	 & 0.75 &  24  &  22	   &  27   &  28 $\pm$ 6  & 151 $\pm$ 5\\
6195.97   &   6220.17	& 1170.9   $\pm$ 4.8	&  15 $\pm$ 2	 & 0.37 &   11  &   20	   &  10   &  10 $\pm$ 3  &   30 $\pm$ 15\\
6203.08$^b$&  6227.28	& 1169.6   $\pm$ 9.7	&  26 $\pm$ 3	 & 1.35 &  19  &\multirow{2}*{\resizebox{5mm}{5mm}{\Large \} } } \multirow{2}*[0mm]{66}	   &  11   &  50 $\pm$ 20 & 191 $\pm$ 5\\
6204.66$^b$&  6228.50	& 1151.9  $\pm$ 9.7   &  76 $\pm$ 4	 & 4.6 &  34  &  	   &  18   &\multicolumn{2}{c}{included in 6203}\\
6269.75   &   6294.15	& 1166.7   $\pm$ 14.3   &  35 $\pm$ 8	 & 1.65 &  14  &  23	   &  10   &   4 $\pm$ 8  &	--      \\
6283.85$^g$ &   6308.25	& 1164.1   $\pm$ 23.8	& 500 $\pm$ 80   & 2.5 &  111  &  390	   &  111   & 225 $\pm$ 21 & --	      \\
6379.29   &   6404.19	& 1170.2   $\pm$ 4.5	&  12 $\pm$ 3	 & 0.48 &  14  &  14	   &  24   &  55 $\pm$ 14  &  75 $\pm$ 8\\   
6613.56   &   6639.36	& 1169.5   $\pm$ 4.1	&  52 $\pm$ 3	 & 1.00 &  42  &  40	   & 43   &  19 $\pm$ 6  &	--      \\
6660.64   &   6686.77	& 1176.1   $\pm$ 4.6	&  13 $\pm$ 5	 & 0.70 &   9  &   --	   &  --   &    --	    &	--      \\  
6993.18   &   7020.58	& 1166.9   $\pm$ 4.9	&  23 $\pm$ 7	 & 0.79 &  21  &  -- 	   &  --   &    --	    &	--      \\
7223.96   &   7251.96	& 1162.0   $\pm$ 4.1	&  74 $\pm$ 5	 & 0.90 &  47  &  --	   &  --   &    --	    &	--      \\    
\hline
SN\,2003hn: &&&&&&&&&\\          
\hline
5780.37   &  5805.3      & 1293.0 $\pm$ 40       &  52 $\pm$ 7     & 2.4  & & & & &\\ 
6283.85   &  6311.9      & 1338.2 $\pm$ 52       & 130 $\pm$ 11    & 5.0  & & & & &\\

&&&&&&&&&\\
\hline	
\end{tabular} \\
\\
\begin{tabular}{lp{16.5cm}}
$^a$ & Included are DIBs with A$_{c} \gtrsim 0.07$ from the table of Herbig (1995).  
Rest wavelengths from the {\cite{Gala00}} survey. \\
$^b$ & The 6203.10 and 6204.27 DIBs are two partly overlapping DIBs, which are sometimes taken to be a single DIB feature. \\
$^c$ & DIB equivalent width for the MW ``average diffuse cloud'' (Jenniskens \& Desert 1994) scaled to $E(B-V)=0.18$ mag, i.e. that within the host galaxy towards SN\,2001el.\\
$^d$  & Galactic lines of sight with $E(B-V)=$ 0.20 and 0.32 mag for HD\,144217 and HD\,149757, respectively.
	Data from FEROS program 64.H-0224 obtained by one of us (LK).\\
$^e$ & Values for Sk-69 223, $E(B-V) \approx 0.35$ mag, from Cox et al. (in preparation). \\
$^f$ & From D'Odorico et al. (1989), 
Note revised $E(B-V)=0.6$ mag (\cite{Nugent02}).
The equivalent width given for $\lambda$~6203 includes also the $\lambda$~6204 DIB, and the $\lambda$~6379 DIB equivalent width 
includes the $\lambda$~6376 DIB. \\
$^g$ & The FWHM applies to the strong ``narrow'' component of the 6284~\AA\ DIB. The EW includes the broader underlying component.\\ 
\end{tabular}
\label{tb:DIB}
\end{table*}

\begin{table*}
\centering
\caption{ 
The column densities towards SNe 2001el and 2003hn derived by different methods.
The upper panel values are derived from the EW and an optically thin approximation ignoring the effects of saturation
(see text). The lowermost panel shows instead the values derived via the 
doublet ratio (DR) technique (i.e., curve of growth technique applied to doublet lines), 
as well as the total column densities
derived with VPFIT. 
The optically thin
approximation is not valid for the saturated doublet lines
towards NGC\,1448, giving values that underestimate the true column
density.  
}
	 
\begin{tabular}{llll|ll}\hline\hline
\multicolumn{6}{c}{Supernovae in NGC\,1448: Interstellar Atomic Lines} \\ 
    &&\multicolumn{2}{c|}{SN\,2001el}&\multicolumn{2}{c}{SN\,2003hn}\\ \hline
Line	       &     & EW     & log $N$  & EW     & log $N$  \\ 
		&    & (m\AA) & (cm$^{-2}$) & (m\AA) & (cm$^{-2}$)\\ \hline
\ion{Ca}{ii}~K & MW  & 104(1) & 12.04 & 104(4)    & 12.04   \\
	       & NGC & 366(2) & 12.59 & 525(6) & 12.75  \\
&&&&&\\
\ion{Ca}{ii}~H & MW  &  54(1) & 12.06 &  60(6) & 12.10  \\
	       & NGC & 220(1) & 12.67 & 320(7) & 12.83  \\
&&&&&\\
\ion{Na}{i}~D2 & MW  &  37(1) & 11.27 &  35(4) & 11.25  \\
	       & NGC & 367(2) & 12.27 & 503(5) & 12.41 \\
&&&&&\\
\ion{Na}{i}~D1 & MW  &  --    & --     & --	& --	 \\
	       & NGC & 302(2) & 12.49 & 289(7) & 12.47  \\
\hline
\multicolumn{6}{c}{}\\ \hline\hline
&& \multicolumn{4}{c}{column density log $N$ (cm$^{-2}$)}\\ \hline
Line		 &       & DR	 & VPFIT 	     & DR	 & VPFIT  \\ \hline
&&&&&\\
\ion{Ca}{ii} 	& MW 	& 12.07$^{+0.02}_{-0.02}$ & 12.12 $\pm$ 0.05 & 12.17$^{+0.12}_{-0.11}$ & 12.12 $\pm$ 0.14 \\
&&&&&\\
		 & NGC	 & 12.76$^{+0.04}_{-0.05}$ & 12.79 $\pm$ 0.08 & 12.93$^{+0.17}_{-0.18}$	& 12.89  $\pm$ 0.20 \\
&&&&&\\
\ion{Na}{i} 	& NGC	& 12.88$^{+0.03}_{-0.03}$ & 12.76 $\pm$ 0.04& 12.52$^{+0.03}_{-0.02}$ 	& 12.55	$\pm$ 0.09 \\
&&&&&\\
\hline
\end{tabular}
\label{tb:EWs}
\end{table*}

\clearpage

\end{document}